\documentclass[12pt]{aastex63}

\newcommand {\kms }{$\,$km s$^{-1}$}
\definecolor{DrakGreen}{HTML}{006400}
\definecolor{LimeBlue}{HTML}{008b8b}
\accepted{June 11, 2023, for publication in the ApJ.}
\shorttitle{OH Outflow in G34.26+0.15}
\shortauthors{Tan et al.}
\graphicspath{{./}{figures/}}
\usepackage{subfigure}

\begin{document}

\title{Excited Hydroxyl Outflow in the High-Mass Star-Forming Region G34.26+0.15}

\correspondingauthor{Esteban D. Araya}
\email{ed-araya@wiu.edu}

\author{Wei Siang Tan}
\affiliation{Physics Department, Western Illinois University, 
1 University Circle, Macomb, IL 61455, USA.}
\affiliation{National University of Singapore, 21 Lower Kent Ridge Rd, Singapore 119077.}

\author[0000-0001-6755-9106]{Esteban D. Araya}
\affiliation{Physics Department, Western Illinois University, 
1 University Circle, Macomb, IL 61455, USA.}
\affiliation{New Mexico Institute of Mining and Technology, Physics Department, 801 Leroy Place, Socorro, NM 87801, USA.}

\author{Cade Rigg}
\affiliation{Southeastern High School, 90 W Green St, Augusta, IL 62311, USA.}

\author{Peter Hofner}
\altaffiliation{Adjunct Astronomer at the National Radio Astronomy Observatory, 1003 Lopezville Road, Socorro, NM 87801, USA.}
\affiliation{New Mexico Institute of Mining and Technology, Physics Department, 801 Leroy Place, Socorro, NM 87801, USA.}

\author{Stan Kurtz} 
\affiliation{Instituto de Radioastronom\'{i}a y Astrof\'{i}sica,
Universidad Nacional Aut\'{o}noma de M\'{e}xico, Apdo. Postal 3-72, 
58089, Morelia, Michoac\'{a}n, Mexico.}

\author{Hendrik Linz}
\affiliation{Max Planck Institute for Astronomy,  K{\"o}nigstuhl 17, 69117, Heidelberg, Germany.}

\author{Viviana Rosero}
\affiliation{National Radio Astronomy Observatory, 1003 Lopezville Road, Socorro, NM 87801, USA.}



\begin{abstract}

G34.26+0.15 is a region of high-mass star formation that contains a broad range of young stellar objects in different stages of evolution, including a hot molecular core, hyper-compact H{\small II} regions and a prototypical cometary ultra-compact H{\small II} region. Previous high-sensitivity single dish observations by our group resulted in the detection of broad 6035$\,$MHz OH absorption in this region; the line showed a significant blue-shifted asymmetry indicative of molecular gas expansion. We present high-sensitivity Karl G. Jansky Very Large Array (VLA) observations of the 6035$\,$MHz OH line conducted to image the absorption and investigate its origin with respect to the different star formation sites in the region. In addition, we report detection of 6030$\,$MHz OH absorption with the VLA and further observations of 4.7$\,$GHz and 6.0$\,$GHz OH lines obtained with the Arecibo Telescope. The 6030$\,$MHz OH line shows a very similar absorption profile as the 6035$\,$MHz OH line. We found that the 6035$\,$MHz OH line absorption region is spatially unresolved at $\sim 2$\arcsec~scales, and it is coincident with one of the bright ionized cores of the cometary H{\small II} region that shows broad radio recombination line emission. We discuss a scenario where the OH absorption is tracing the remnants of a pole-on molecular outflow that is being ionized inside-out by the ultra-compact H{\small II} region.

\end{abstract}



\section{Introduction} \label{Intro}

In the turbulent core accretion model of high-mass star formation, massive stars ($M \gtrsim 8\,$M$_{\odot}$) form in Giant Molecular Clouds, where gravitational collapse occurs in dense cores leading to the formation of a protostar surrounded by a disk through which accretion takes place (e.g.,  \citealt{Motte_2018ARA&A..56...41M},  \citealt{Sanchez-Monge_2013A&A...552L..10S},  \citealt{Zinnecker_2007ARA&A..45..481Z}, \citealt{Beuther_and_Shepherd_2005ASSL..324..105B}). In this phase, ionized jets and molecular outflows form (e.g., \citealt{Rosero_2019ApJ...880...99R}, \citealt{Kuiper_2018A&A...616A.101K}) while the protostar continues increasing mass, until thermonuclear fusion begins and high energy UV photons start to photoionize the natal environment. It is thought that this process leads to the formation of hypercompact H{\small II} (HCH{\small II}) and ultracompact H{\small II} (UCH{\small II}) regions (e.g., \citealt{Yang_2021A&A...645A.110Y}) and eventually classical H{\small II} regions. During the early phases of high-mass star formation, before the formation and during the early evolution of UCH{\small II} regions, hot molecular cores (HMC) develop, which are among the most chemically rich environments in the interstellar medium (ISM; e.g., \citealt{Kurtz_2000prpl.conf..299K}; \citealt{Cesaroni_2005IAUS..227...59C}; \citealt{van_der_Tak_2005IAUS..227...70V}). Although most HMC are thought to be centrally heated, others appear to be externally heated by nearby UCH{\small II} regions \citep{Mookerjea_Kinematics_and_chemistry_2007ApJ...659..447M}. Despite the great progress in our understanding of high-mass star formation in the last few decades, key aspects remain unclear, such as the evolution of the outflows/jets from their initial launching to their later interaction with UCH{\small II} regions. UCH{\small II} and HCH{\small II} regions characterized by broad radio recombination lines (RRLs) could be key to reveal the evolution of outflows after photoionization begins to dominate (e.g., \citealt{Sewilo_2004ApJ...605..285S}).

\subsection{G34.26+0.15: A high-mass star-forming region with OH Absorption}\label{Intro_G34}

At a distance of approximately 3.0$\,$kpc\footnote{We adopt the trigonommetric parallax distance of 3.0$\,$kpc of G034.43+0.24 \citep{Mai_2023arXiv230309129M} as the distance to G34.26+0.15. A 3$\,$kpc kinematic distance has also been reported for G34.26+0.15 (e.g., \citealt{Araya_distance_g34_2002ApJS..138...63A}, \citealt{Kuchar_distance_1994ApJ...436..117K}). However, we note that \cite{Urquhart_2018MNRAS.473.1059U} listed a closer distance.},
the high-mass star forming region G34.26+0.15 is an excellent example where all aspects of high-mass star formation coexist. G34.26+0.15 is part of a large scale ($\sim 30\,$pc, i.e., 34\arcmin~at 3.0$\,$kpc) star forming filament, including the Infrared Dark Cloud IRDC G34.43+0.24 \citep{Xu_2016ApJ...819..117X}. G34.26+0.15 contains H{\small II} regions at different stages of evolution, including two HCH{\small II} regions (see resolved images of the HCH{\small II} regions in \citealt{Avalos_2009ApJ...690.1084A}), and a HMC that appears to be externally heated by a cometary UCH{\small II} region \citep{Mookerjea_Kinematics_and_chemistry_2007ApJ...659..447M}. Infall signatures of molecular gas have been detected in G34.26+0.15 (\citealt{Wyrowski_2012A&A...542L..15W}, \citealt{Liu_2013ApJ...776...29L},  \citealt{Wyrowski_2016A&A...585A.149W}; \citealt{Hajigholi_accretion_process_2016A&A...585A.158H}), and outflow activity is evident by the detection of SiO emission \citep{Hatchell_2001A&A...372..281H}, and a cone-like 4.5$\,\mu$m structure indicative of shocked gas (\citealt{Sewilo_HII2011ApJS..194...44S}, \citealt{Liu_2013ApJ...776...29L}). The cometary UCH{\small II} G34.26+0.15 is one of the best examples of ionized gas with broad RRL emission (e.g., \citealt{Sewilo_2004ApJ...605..285S}; \citealt{Rodriguez_Rico_2002RMxAA..38....3R}; \citealt{Gaume_1994ApJ...432..648G}).

As it is typical of active sites of high-mass star formation, many molecular species have been detected as masers in G34.26+0.15, including H$_2$O (e.g., \citealt{Hofner_1996A&AS..120..283H}, \citealt{Imai_2011PASJ...63.1293I}), NH$_3$
\citep{Yan_2022A&A...659A...5Y}, CH$_3$OH
 (e.g., \citealt{Bartkiewicz_2016A&A...587A.104B}, \citealt{Kurtz_2004ApJS..155..149K}) and OH 
(e.g., \citealt{Garay_1985ApJ...289..681G},  \citealt{Tan_arecibo_10.1093/mnras/staa1841}). In the case of hydroxyl, masers of the 18$\,$cm ground state have been detected (e.g., \citealt{Beuther_2019A&A...628A..90B}, \citealt{Gasiprong_2002MNRAS.336...47G}, \citealt{Zheng_2000MNRAS.317..192Z}), but to our knowledge, no absorption has been found \citep{Rugel_thor_abundance_2018A&A...618A.159R}. In contrast, using the 305$\,$m Arecibo Telescope, \cite{Al-Marzouk_2012ApJ...750..170A} reported the detection of 6035$\,$MHz OH masers as well as a relatively weak ($-65\,$mJy) and broad ($FWHM = 18$\kms) absorption line (6035$\,$MHz OH masers were also detected by \citealt{Caswell_1995MNRAS.273..328C} and \citealt{Caswell_2001MNRAS.326..805C}, but no absorption was reported). The absorption was confirmed by \cite{Tan_arecibo_10.1093/mnras/staa1841}\footnote{Additional observations of the absorption line were presented in the MS thesis of Kim (2014; Western Illinois University).}, where we reported a highly asymmetric blue-shifted 6035$\,$MHz OH profile. The line has a velocity linewidth at zero intensity greater than $\sim 50$\kms, indicative of outflowing molecular gas. The 6035$\,$MHz line belongs to the 6.0$\,$GHz OH ladder (6016, 6030, 6035, 6049$\,$MHz) that corresponds to the first excited rotational level ($J=5/2$) of the $2\Pi_{3/2}$ ladder above the well known ground state 18$\,$cm OH transitions; the 6.0$\,$GHz transitions are at energy levels $E_{lower}/{k_B} = 120\,$K\footnote{Spectroscopy information of all transitions in this article is from Splatalogue (\url{https://splatalogue.online/}).} (e.g., see energy level diagrams in \citealt{Gray_2012msa..book.....G}, \citealt{Pihlstrom_2008ApJ...676..371P}, \citealt{Desmurs_2002A&A...394..975D}). 
Absorption of the 6035 and 6030$\,$MHz lines has been detected in the ISM, including toward other high-mass star forming regions (e.g., \citealt{Tan_arecibo_10.1093/mnras/staa1841}, \citealt{Avison_2016MNRAS.461..136A}, \citealt{Fish_2006A&A...458..485F}, \citealt{Baudry_1997A&A...325..255B}, \citealt{Guilloteau_1984A&A...131...45G}).

In this work we present a follow up study of the broad 6035$\,$MHz OH absorption line in G34.26+0.15 based on NSF's Karl G. Jansky Very Large Array (VLA) and 305$\,$m Arecibo Telescope data, including observations of other OH transitions of the 6.0 and 4.7$\,$GHz states. The 4.7$\,$GHz OH transitions (4660, 4765, and 4751$\,$MHz) correspond to the lowest state of the $2\Pi_{1/2}$ ladder ($J=1/2$; $E_{lower}/k_B = 181\,$K; \citealt{Gray_2012msa..book.....G}); transitions from this state have also been detected toward star forming regions and late type stars (e.g., \citealt{Qiao_2022ApJ...928..129Q}, \citealt{Strack_2019ApJ...878...90S}, \citealt{Palmer_2005MNRAS.360..993P},
\citealt{Dodson_2002MNRAS.333..307D},
\citealt{Cohen_1991MNRAS.250..611C}). In Section~\ref{Obs} we report details of the observations and data reduction; results and discussion are presented in sections~\ref{Results} and \ref{Analysis_and_Disc}, respectively. We summarize our work in Section~\ref{Summary}.

\newpage

\section{Observations} \label{Obs}

\subsection{Arecibo Telescope}\label{Obs:Arecibo}

Observations of the 4.7$\,$GHz OH (4660.2420$\,$MHz, 4750.6560$\,$MHz, 4765.5620$\,$MHz) and 6.0$\,$GHz OH transitions (6016.74600$\,$MHz, 6030.74850$\,$MHz, 6035.09320$\,$MHz, 6049.08400$\,$MHz) were conducted with the 305$\,$m Arecibo Telescope in Puerto Rico, between October 2008 and January 2010 as part of the project A2411, which was designed to investigate variability of OH masers. The maser variability aspect of the observations was reported in \cite{Al-Marzouk_2012ApJ...750..170A}. The observations were conducted in position switching (1 or 2$\,$min ON-source followed by OFF-source observations for the same amount of time, and 10 second calibration observations by injecting noise from a diode). The WAPP (Wideband Arecibo Pulsar Processor) spectrometer was used for the observations, with a bandwidth of 3.125$\,$MHz and 2048 channels per transition (Table~\ref{tab:oh_obs}). The lines were observed in two groups, the 4.7$\,$GHz lines using the C-Band receiver, and the 6.0$\,$GHz transitions using the C-Band High receiver. Most scans included observations of the Arecibo Telescope calibrator B1857+129, which resulted in typical system temperatures between 25 and 33$\,$K, and telescope gains of approximately 8 and 6$\,$K$\,$Jy$^{-1}$ at C-band and C-Band High frequencies, respectively; pointing errors smaller than 17\arcsec, and half-power beam widths of 56\arcsec at 4860$\,$MHz and 44\arcsec at 6600$\,$MHz.

A total of 23 scans were observed of the 4.7$\,$GHz transitions and 16 scans of the 6.0$\,$GHz lines. Data reduction and calibration was done in IDL using the AOIDL library provided by the observatory. After checking for consistency between polarization and different scans, and after excluding scans affected by radio frequency interference and technical problems (such as high system temperatures due to a cooling failure of the C-Band High receiver), we subtracted linear baselines obtained from line-free channels at both sides of the spectral lines, averaged the scans, and smoothed them to final channel widths between 1 and 8\kms~to enhance signal-to-noise of weak broad lines. In the case of the 6035$\,$MHz OH lines, we smoothed the data to a channel width of $1$\kms~to avoid spectral contamination of the peak absorption with narrow bright masers. A discussion of the 6035$\,$MHz OH masers was reported in \cite{Al-Marzouk_2012ApJ...750..170A}; in this paper we concentrate on the broad absorption lines. Further details of system checking observations, calibration and data reduction are included in \cite{Al-Marzouk_2012ApJ...750..170A}.

\subsection{Very Large Array (VLA)}\label{Obs:VLA}

The 6.0$\,$GHz OH transitions observed with the Arecibo Telescope were also observed with the VLA in two configurations: D-Array (September 2018; project ID 18A$-$401) and B-Array (June 2020; project ID 20A$-$311). We used the quasar 3C286 as flux and bandpass calibrator. J1856+0610 and J1824+1044 were used as complex gain (amplitude and phase) calibrators for the B and D configuration observations, respectively. All lines were simultaneously observed in different spectral windows. Further details of the VLA observations are reported in Table~\ref{tab:oh_obs}.

Data reduction was done using the Common Astronomy Software Applications package (CASA\footnote{https://casa.nrao.edu}). The VLA pipeline in CASA 5.4.1-31 (pipeline 51-P2-B)\footnote{https://science.nrao.edu/facilities/vla/data-processing/pipeline} was used for calibration of the VLA D-configuration data. The VLA pipeline in CASA 5.6.2-3 (pipeline 56-P2-B) was used to calibrate the VLA B-configuration observations. Continuum subtraction was done in the $(u,v)$ plane using the task {\tt uvcontsub}. Data cubes were obtained with the task {\tt tclean} using {\it natural} weighting. We obtained velocity moments (integrated intensity, velocity, and velocity dispersion) using the task {\tt immoments}. The cubes were smoothed to final channel widths between 0.2 and 1.5\kms~to optimize the detection of weak OH absorption while minimizing spectral contamination by the narrow and bright masers at 6035$\,$MHz.

\newpage

\begin{table*}
    \centering
	\caption{Summary of OH Observations.}
	\label{tab:oh_obs}
	\begin{tabular}{lccc} 
		\hline
		\hline
		Telescope                  & Arecibo              & VLA-D              & VLA-B\\
		\hline
		Observing Epoch            & 2008 - 2010          & September 2018     & June 2020\\
		Project ID                 & A2411                & 18A$-$401          & 20A$-$311\\
		Pointing RA (J2000)$^{(1)}$    & 18:53:18.5           & 18:53:18.5 & 18:53:18.5\\
		Pointing Dec (J2000)$^{(1)}$   & +01:14:59            & +01:14:59 & +01:14:59\\
		Configuration              & \nodata              & D                  & B \\
		Bandwidth                  & 3.125$\,$MHz         & 4$\,$MHz           & 4$\,$MHz \\
		Number of Channels         & 2048                 & 2048               & 2048 \\
		Channel Separation         & 1.53$\,$kHz          & 1.95$\,$kHz        & 1.95$\,$kHz \\
		Beamwidth$^{(2)}$          & 44\arcsec, 56\arcsec &19.7\arcsec$\times$9.7\arcsec, $-$19$^{\circ}$  & 2.2\arcsec$\times$1.3\arcsec, $-$36$^{\circ}$ \\
        Rest Frequencies$^{(3)}$   & 4660.2420$\,$MHz     & \nodata            & \nodata \\
                                   & 4750.6560$\,$MHz     & \nodata            & \nodata \\
                                   & 4765.5620$\,$MHz     & \nodata            & \nodata \\
                                   & 6016.7460$\,$MHz     & 6016.7460$\,$MHz   & 6016.7460$\,$MHz \\ 
                                   & 6030.7485$\,$MHz     & 6030.7485$\,$MHz   & \nodata \\
                                   & 6035.0932$\,$MHz     & 6035.0932$\,$MHz   & 6035.0932$\,$MHz \\
                                   & 6049.0840$\,$MHz     & 6049.0840$\,$MHz   & 6049.0840$\,$MHz \\
		\hline
		\multicolumn{4}{p{14cm}}{$^{(1)}$ G34.26+0.15 pointing coordinates.}\\	
		\multicolumn{4}{p{14cm}}{$^{(2)}$ The half-power beam width at 6600$\,$MHz and 4860$\,$MHz are reported for the Arecibo observations. Synthesized beam at 6035$\,$MHz 
		($\theta_{maj} \times \theta_{min}$, PA) is reported for the VLA observations. }\\
  		\multicolumn{4}{p{14cm}}{$^{(3)}$ Rest frequencies of OH transitions with reliable detection or non-detection. Spectroscopy information is from Splatalogue (https://splatalogue.online/).}
		\end{tabular}
\end{table*}

\section{Results}\label{Results}

\subsection{Arecibo} \label{Res:OH-Arecibo}

\begin{center}
\begin{figure}
\centering
\includegraphics[width=16cm]{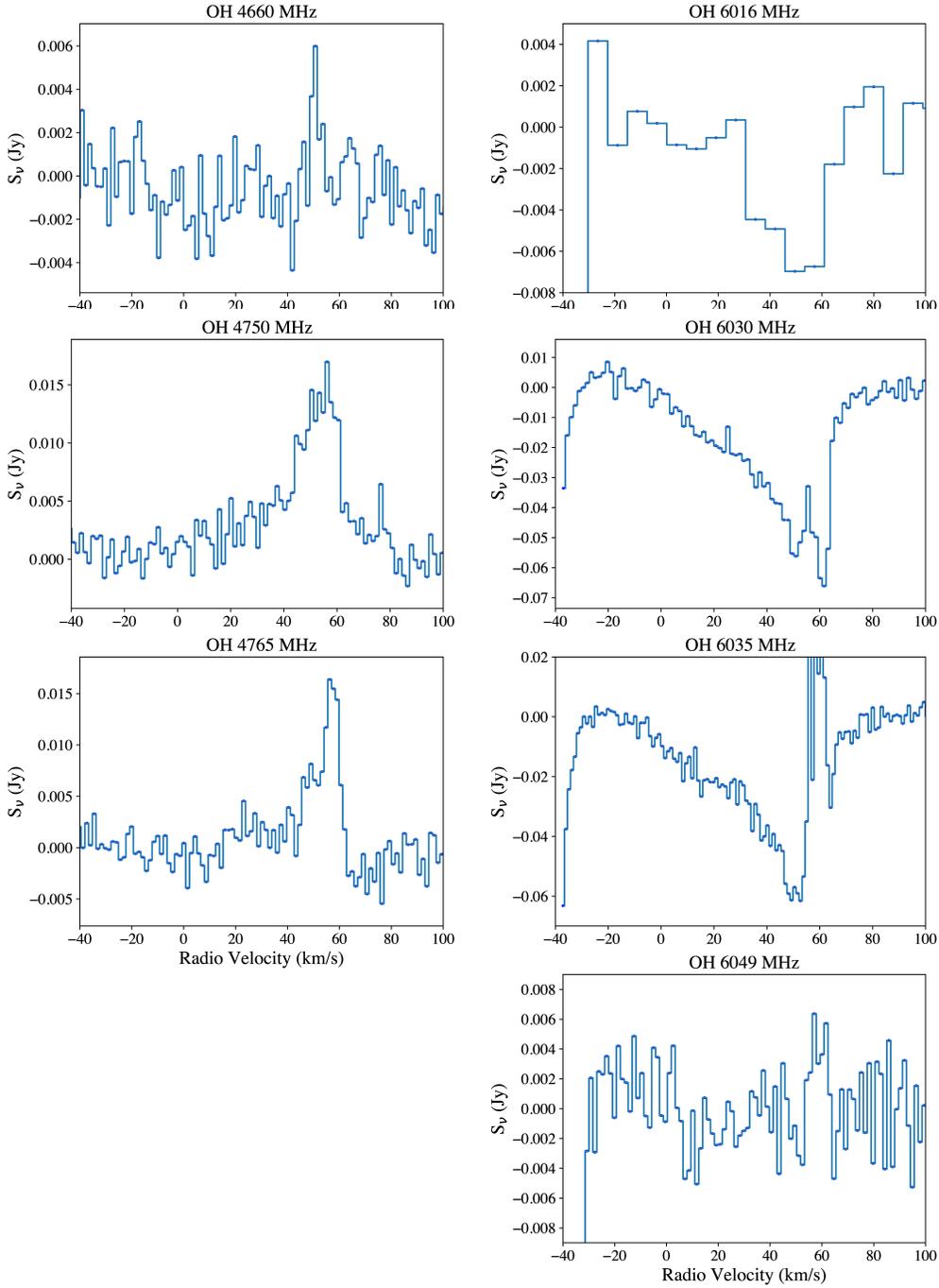}\\
\caption{OH spectra obtained with the Arecibo Telescope. The left panels show the 4.7$\,$GHz OH transitions and the right panels show the 6.0$\,$GHz lines (the end of the bandpass can be seen at $-$40\kms~in the 6.0$\,$GHz spectra). We report detection of emission in the 4750$\,$MHz and 4765$\,$MHz transitions (the 4660$\,$MHz may also show a narrow and weak emission line, however, only one channel is above 3$\sigma$). Maser lines are detected in the 6035$\,$MHz transition (see \citealt{Al-Marzouk_2012ApJ...750..170A}), and absorption is found in the 6035$\,$MHz, 6030$\,$MHz and 6016$\,$MHz transitions. The 6016$\,$MHz line is weak; the spectra had to be smoothed to a channel width of 7.6\kms~to detect the absorption. We report no detection of the 6049$\,$MHz transition (no clear evidence of absorption is detected when the spectrum is smoothed).}
\label{fig:Arecibo_spectra}
\end{figure}
\end{center}

As reported in \cite{Al-Marzouk_2012ApJ...750..170A}, bright and narrow OH masers were detected in the 6035$\,$MHz transition, and a weak and broad absorption feature was found when all scans were averaged. As shown in Figure~\ref{fig:Arecibo_spectra} and Table~\ref{tab:OH_Line_Parameters}, absorption was also detected in the 6030$\,$MHz OH transition; broad and weak absorption was also detected in the 6016$\,$MHz OH transition. We found no evidence of absorption in the 6049$\,$MHz OH line. 

In contrast to the 6.0$\,$GHz OH transitions that show broad absorption, no absorption was detected in the 4.7$\,$GHz OH transitions, instead emission was found in the 4750$\,$MHz and 4765$\,$MHz OH lines, and a tentative (weak and narrow) emission line was found in the 4660$\,$MHz OH spectrum (Table~\ref{tab:OH_Line_Parameters}, Figure~\ref{fig:Arecibo_spectra}). In the case of the 4750$\,$MHz and 4765$\,$MHz OH detections, the lines have asymmetric profiles, suggesting overlapping velocity components. 
Emission of these transitions (including broad and narrow lines) has been detected toward other high-mass star forming regions (e.g., \citealt{Qiao_2022ApJ...928..129Q}, \citealt{Cohen_1991MNRAS.250..611C}, and references therein; note that \citealt{Qiao_2022ApJ...928..129Q} did not detect the 4.7$\,$GHz OH lines toward G34.26+0.15 due to sensitivity). The low RMS obtained in the short observations reported in this work (RMS $\lesssim 3 \,$mJy in $\approx 20\,$min ON-source) highlights the extraordinary sensitivity that the 305$\,$m Arecibo Telescope was able to achieve, and the great loss in high-sensitivity spectroscopic capabilities at frequencies below $\sim 10\,$GHz with the demise of the telescope.

\newpage

\begin{center}
\begin{longtable}{lcccccc}
	\caption{OH Line parameters.} 	\label{tab:OH_Line_Parameters}\\
	\hline
    \multicolumn{1}{c}{{Spectral Line}} & \multicolumn{1}{c}{{S$_\nu$}} & \multicolumn{1}{c}{{RMS}} & \multicolumn{1}{c}{{V$_{LSR}$}} & \multicolumn{1}{c}{{Width}} & \multicolumn{1}{c}{{$\int$S$_\nu$dv}} & \multicolumn{1}{c}{{Telescope}}\\ 
    \multicolumn{1}{c}{{}} & \multicolumn{1}{c}{{(Jy)}} & \multicolumn{1}{c}{{(Jy)}} & \multicolumn{1}{c}{{(km$\,$s$^{-1}$)}} & \multicolumn{1}{c}{{(km$\,$s$^{-1}$)}} & \multicolumn{1}{c}{{(Jy$\,$km s$^{-1}$})} & \\
    \hline 
    \endfirsthead
	\multicolumn{7}{c}%
    {{\tablename\ \thetable{} -- continued from previous page}} \\
    \hline
    \multicolumn{1}{c}{{Spectral Line}} & \multicolumn{1}{c}{{S$_\nu$}} & \multicolumn{1}{c}{{RMS}} & \multicolumn{1}{c}{{V$_{LSR}$}} & \multicolumn{1}{c}{{Width}} & \multicolumn{1}{c}{{$\int$S$_\nu$dv}} & \multicolumn{1}{c}{{Telescope}}\\ 
    \multicolumn{1}{c}{{}} & \multicolumn{1}{c}{{(Jy)}} & \multicolumn{1}{c}{{(Jy)}} & \multicolumn{1}{c}{{(km$\,$s$^{-1}$)}} & \multicolumn{1}{c}{{(km$\,$s$^{-1}$)}} & \multicolumn{1}{c}{{(Jy$\,$km s$^{-1}$)}} & \\
    \hline
    \endhead
        4660$\,$MHz     & ...    & 0.0015 & ...       & ...       &...              & Arecibo (1)  \\
        4750$\,$MHz     & 0.017   & 0.0013 & 56.3(1.4) & 30.3(2.9) & 0.28(0.03)      & Arecibo \\
        4765$\,$MHz     & 0.016   & 0.002 & 56.2(1.4) & 15.8(2.9) & 0.15(0.03)      & Arecibo \\
		6016$\,$MHz     &$-$0.0070& 0.0007& 49.5(7.6) & 38.0(15.2)& $-$0.19(0.04)   & Arecibo \\
		                &...      & 0.008  & ...       &...        &...              & VLA$-$D\\
		                &...      & 0.009  & ...       &...        &...              & VLA$-$B\\
        6030$\,$MHz     &$-$0.066 & 0.002  & 61.7(1.5) & 63.7(3.0) & $-$1.89(0.06) & Arecibo \\
                        &$-$0.052 & 0.004  & 62.0(1.5) & 50.0(3.0) & $-$1.48(0.11)   & VLA$-$D\\
        6035$\,$MHz     &$-$0.062 & 0.003  & 52.7(1.1) & $>$55     &$>|-1.6|$          & Arecibo (2)\\
                        &$-$0.056 & 0.005  & 51.7(1.5) & $>$49     &$>|-1.5|$          & VLA$-$D (3)\\
                        &0.104 &0.012 &62.4(0.1) &0.5(0.2) &0.03(0.01) & VLA$-$D\\
                        &0.091 &0.012 &62.0(0.1) &0.4(0.2) &0.02(0.01) & VLA$-$D\\
                        &0.246 &0.012 &60.9(0.1) &0.9(0.2) &0.11(0.01) & VLA$-$D\\
                        &0.091 &0.012 &60.1(0.1) &0.5(0.2) &0.03(0.01) & VLA$-$D\\
                        &0.287 &0.012 &59.5(0.1) &0.7(0.2) &0.10(0.01) & VLA$-$D\\
                        &0.496 &0.012 &58.5(0.1) &0.7(0.2) &0.20(0.01) & VLA$-$D\\
                        &0.413 &0.012 &58.1(0.1) &0.7(0.2) &0.20(0.01) & VLA$-$D\\
                        &0.094 &0.012 &56.8(0.1) &0.6(0.2) &0.04(0.01) & VLA$-$D\\
                        &0.152 &0.012 &55.7(0.1) &0.7(0.2) &0.06(0.01) & VLA$-$D\\
                        &$-$0.058 &0.003 &50.8(1.5) &58.2(3.0) &$-$1.86(0.08)&VLA$-$B\\
                        &0.115 &0.008 &62.4(0.1) &1.1(0.2) &0.08(0.01) & VLA$-$B\\
                        &0.187 &0.008 &60.9(0.1) &1.2(0.2) &0.14(0.01) & VLA$-$B\\
                        &0.237 &0.008 &59.6(0.1) &1.0(0.2) &0.13(0.01) & VLA$-$B\\
                        &0.481 &0.008 &58.4(0.1) &1.7(0.2) &0.43(0.01) & VLA$-$B\\
                        &0.118 &0.008 &56.8(0.1) &0.9(0.2) &0.08(0.01) & VLA$-$B\\
                        &0.140 &0.008 &55.8(0.1) &0.9(0.2) &0.08(0.01) & VLA$-$B\\
        6049$\,$MHz     & ...     & 0.003  & ...        & ...      & ...             & Arecibo \\   
                        & ...     & 0.008  & ...        & ...      & ...             & VLA$-$D\\
                        & ...     & 0.010  & ...        & ...      & ...             & VLA$-$B\\
		\hline
\multicolumn{7}{p{12cm}}{ We report the channel width as uncertainty in the $V_{LSR}$ peak velocity, and twice the channel width as uncertainty in the linewidth. The linewidth is the velocity range of the line above 3$\sigma$. The uncertainty in the integrated flux density is the 3 $\times$ RMS $\times$ channel width $\times$ square root of the number of channels in the line. }\\ 
\multicolumn{7}{p{12cm}}{(1) Tentative line detected, only one channel above 3$\sigma$, $S_\nu = 6.0\,$mJy, $V_{LSR} = 50.7$\kms.}\\	
\multicolumn{7}{p{12cm}}{(2) Multiple maser peaks overlap the absorption feature, therefore, the linewidth and integrated flux density are lower limits. The data were smoothed to improve signal-to-noise detection of the absorption feature, a discussion of the masers from this dataset with high spectral resolution is presented in \cite{Al-Marzouk_2012ApJ...750..170A}. }\\
\multicolumn{7}{p{12cm}}{(3) Multiple maser peaks overlap the absorption feature, therefore, the linewidth and integrated flux density are lower limits. }\\

\end{longtable} 
\end{center}

\newpage
\subsection{Very Large Array}\label{Res:OH-VLA}

\begin{center}
\begin{figure}
\centering
\vspace{-2.5cm}
\includegraphics[width=19cm]{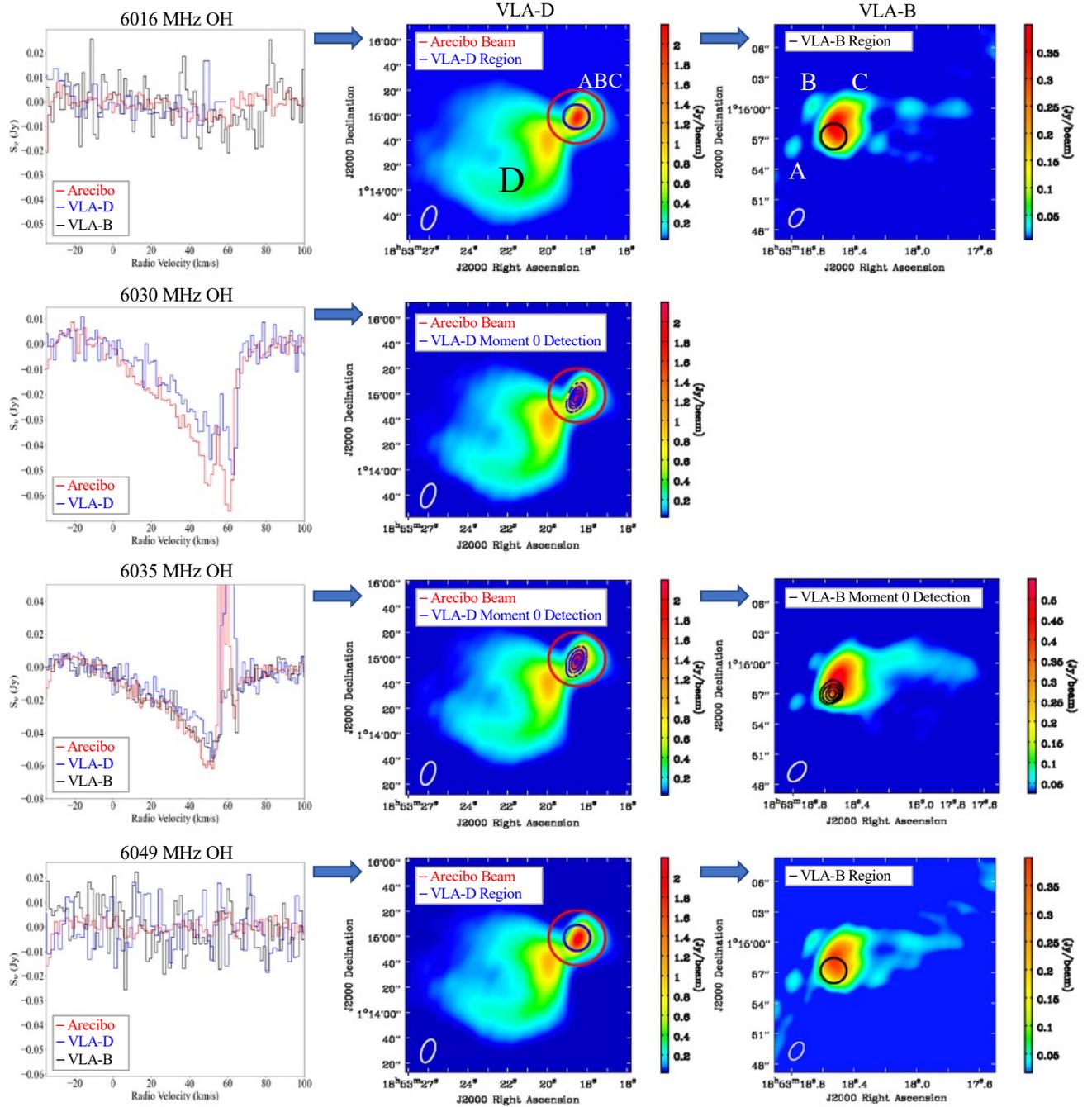}
\vspace{-3.8cm}
\caption{\small Spectra of the 6.0$\,$GHz OH transitions obtained with the Arecibo Telescope (red), VLA in D-configuration (blue) and B-configuration (black) are shown in the left panels; the spectra are sorted by frequency: 6016$\,$MHz, 6030$\,$MHz, 6035$\,$MHz, and 6049$\,$MHz from top to bottom. Radio continuum obtained from the line-free channels of the respective spectral windows is shown in the middle (D-configuration) and right panels (B-configuration). In the middle panels, we show with a red circle the $\sim$44\arcsec~Arecibo beam, centered at the pointing position; the blue contours in the 6030 and 6035$\,$MHz panels show the velocity integrated intensity (moment-0) images of the absorption from the VLA D-configuration cubes (6030$\,$MHz OH absorption contour levels: $-$10, $-$90, $-$170, $-$250 $\times\,$3.08$\,$mJy$\,$km$\,$s$^{-1}$; 6035$\,$MHz OH absorption contour levels: $-$70, $-$180, $-$290, $-$400 $\times\,$2.65$\,$mJy$\,$km$\,$s$^{-1}$). No detection was obtained in the D-configuration observations of the 6016$\,$MHz (middle, upper panel) nor the 6049$\,$MHz (middle, lower panel) OH transitions; for these transitions, we show with a blue circle the region used to obtain the blue spectrum shown in the respective left panels. The right panels are equivalent to the middle panels, but for the B-configuration observations; black circles are the regions integrated to obtain the black spectra shown in the right panels, and the moment-0 image of the 6035$\,$MHz OH absorption is shown in black contours ($-$60, $-$140, $-$220, $-$300 $\times\,$2.19$\,$mJy$\,$km$\,$s$^{-1}$). The spectral window of the 6030$\,$MHz VLA-B observations was corrupted.}
\label{fig:AO-VLA_Spectra_and_Images}
\end{figure}
\end{center}

Figure~\ref{fig:AO-VLA_Spectra_and_Images} summarizes the results obtained with the VLA. The left panels show the 6.0$\,$GHz OH spectra obtained with the VLA in the D-configuration (blue), the spectra obtained in the B-configuration (black) and the Arecibo spectra (red; the 6030, 6035 and 6049$\,$MHz are as in Figure~\ref{fig:Arecibo_spectra}, while the 6016$\,$MHz spectrum is shown with higher spectral resolution than the one shown in Figure~\ref{fig:Arecibo_spectra} to facilitate comparison with the VLA spectra). The center panels show the 6.0$\,$GHz radio continuum from the line-free channels of the respective spectral windows obtained with the VLA in D-configuration, and the right panels show the equivalent 6$\,$cm radio continuum images from the line-free channels of the VLA B-configuration spectral windows.

\begin{figure}
\centering
\includegraphics[width=19cm]{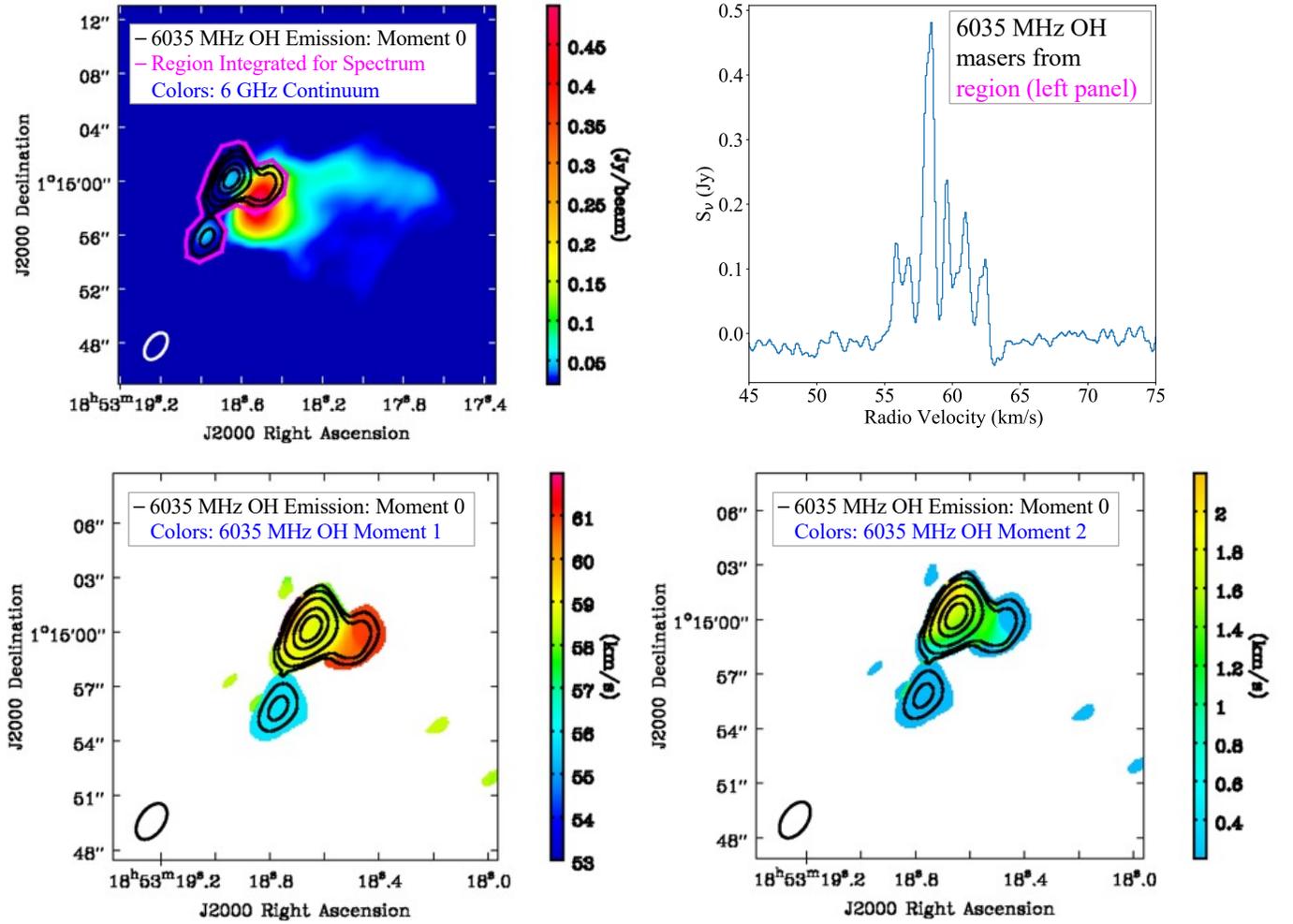}
\caption{The top-left panel shows the radio continuum image obtained from the line-free channels of the VLA B-configuration 6035$\,$MHz OH observations (colors). Contours show the integrated intensity (moment 0) of the 6035$\,$MHz OH masers (contours levels are: 5, 15, 50, 150, 250 $\times\,$2.195$\,$mJy$\,$km$\,$s$^{-1}$). The magenta region represents the area integrated to obtain the spectrum shown in the top-right panel. The flux density and velocity range shown in the spectrum were selected to highlight the masers; the spectrum highlighting the absorption is shown in Figure~\ref{fig:AO-VLA_Spectra_and_Images}. The velocity field (moment 1; colors) and  integrated intensity (moment 0; contours) of the 6035$\,$OH masers are shown in the lower-left panel. The velocity dispersion (moment 2; colors) and integrated intensity (contours) of the 6035$\,$MHz OH masers are shown in the lower-right panel.}
\label{fig:VLA_B_G34_OH_6035_emission MHz}
\end{figure}

As it is evident from Figure~\ref{fig:AO-VLA_Spectra_and_Images}, we obtained almost identical OH absorption profiles of the 6030$\,$MHz and 6035$\,$MHz transitions with Arecibo and the VLA. In the case of the 6035$\,$MHz transition, the Arecibo and VLA D-configuration profiles overlap with strong and narrow OH masers. In contrast, the 6035$\,$MHz OH absorption and masers are spatially separated at the VLA B-configuration resolution. As shown in Figure~\ref{fig:VLA_B_G34_OH_6035_emission MHz}, the masers are located toward the North core of the cometary UCH{\small II} region and toward the two HCH{\small II} regions to the East of the UCH{\small II} region, while the absorption is located toward the southern core of the UCH{\small II} region (Figure~\ref{fig:AO-VLA_Spectra_and_Images}). As this article focuses on the absorption, the masers are not further discussed, nevertheless, the line parameters of the masers are listed in Table~\ref{tab:OH_Line_Parameters} together with the parameters of the absorption lines. Given that the masers are spatially separated from the absorption at the angular resolution of the VLA B-configuration observations, we were able to obtain the absorption spectrum without contamination by OH masers (Figure~\ref{fig:AO-VLA_Spectra_and_Images}). 

Both Arecibo and VLA observations confirm the asymmetric absorption profile of the 6035$\,$MHz OH line reported by \cite{Tan_arecibo_10.1093/mnras/staa1841}. Our data also show that the 6030$\,$MHz OH exhibits a very similar absorption profile, with a blue asymmetry (line-wing) that extends from the peak absorption ($\sim 50 - 60$\kms) to $\sim 0$\kms. The absorption is spatially unresolved (based on a 2D Gaussian fit of the moment-0 VLA-B detection); we obtained consistent line profiles from the angular resolution of the Arecibo Telescope (44\arcsec) to the VLA B-configuration synthesized beam ($\sim 2$\arcsec). To our knowledge, these are the spectrally broadest and most spatially compact 6035 and 6030$\,$GHz OH absorption lines ever detected in the ISM. We note that broad absorption lines ($>100$\kms) of these OH transitions have been detected in other galaxies, but tracing the overall ensemble of molecular clouds seen in absorption against the continuum of the galaxy rather than from an individual site within a star forming region  (\citealt{Eisner_2019ApJ...882...95E}, \citealt{Whiteoak_1990MNRAS.245..665W}).

\clearpage
\section{Analysis and Discussion}\label{Analysis_and_Disc}

\subsection{Origin of the OH Absorption: Association with Broad Radio Recombination Line Ionized Gas}\label{Analysis_and_Disc:Origin}

Based on Arecibo observations of H$_2$CO and the H110$\alpha$ radio recombination line \citep{Araya_distance_g34_2002ApJS..138...63A}, the systemic velocity of the G34.26+0.15 molecular cloud is $\sim$60\kms, and the maximum flux density of the ionized gas emission is at a velocity of 54\kms. As shown in Figures~\ref{fig:Arecibo_spectra} and \ref{fig:AO-VLA_Spectra_and_Images}, the stronger OH absorption detected at 6016$\,$MHz, 6030$\,$MHz and 6035$\,$MHz is approximately within the velocity range traced by H$_2$CO and radio recombination lines. The nature of the broad blue-shifted absorption asymmetry (line-wing) observed in the 6030$\,$MHz and 6035$\,$MHz OH lines is more intriguing. As discussed by \cite{Jacob_CH_2021AA...650..A133}, weak 3$\,$GHz CH emission lines from molecular clouds in the line-of-sight of G34.26+0.15 are detectable at $V_{LSR} \sim 10$ and $\sim 26$\kms, due to molecular clouds in the Sagittarius Spiral Arm (see Figure~2 of \citealt{Jacob_CH_2021AA...650..A133}; the 26\kms~feature was also detected in HCO$^+$ absorption by \citealt{Carral_1992ApJ...385..244C}). We find it unlikely that the broad blue-shifted absorption of the 6030$\,$MHz and 6035$\,$MHz OH lines is caused by molecular clouds unrelated to G34.26+0.15, given the smooth line profile of the absorption from the peak systemic velocity, and the compact nature of the absorption (Figure~\ref{fig:AO-VLA_Spectra_and_Images}). Instead, it is more likely that the blue-shifted line-wing is tracing expansion/outflowing motion of the molecular gas in the region. This is in contrast to ammonia transitions obtained with Herschel that are indicative of infall \citep{Hajigholi_accretion_process_2016A&A...585A.158H}.

\begin{center}
\begin{figure}
\centering
\includegraphics[trim={0 0cm 0 0 },width=0.8\textwidth]{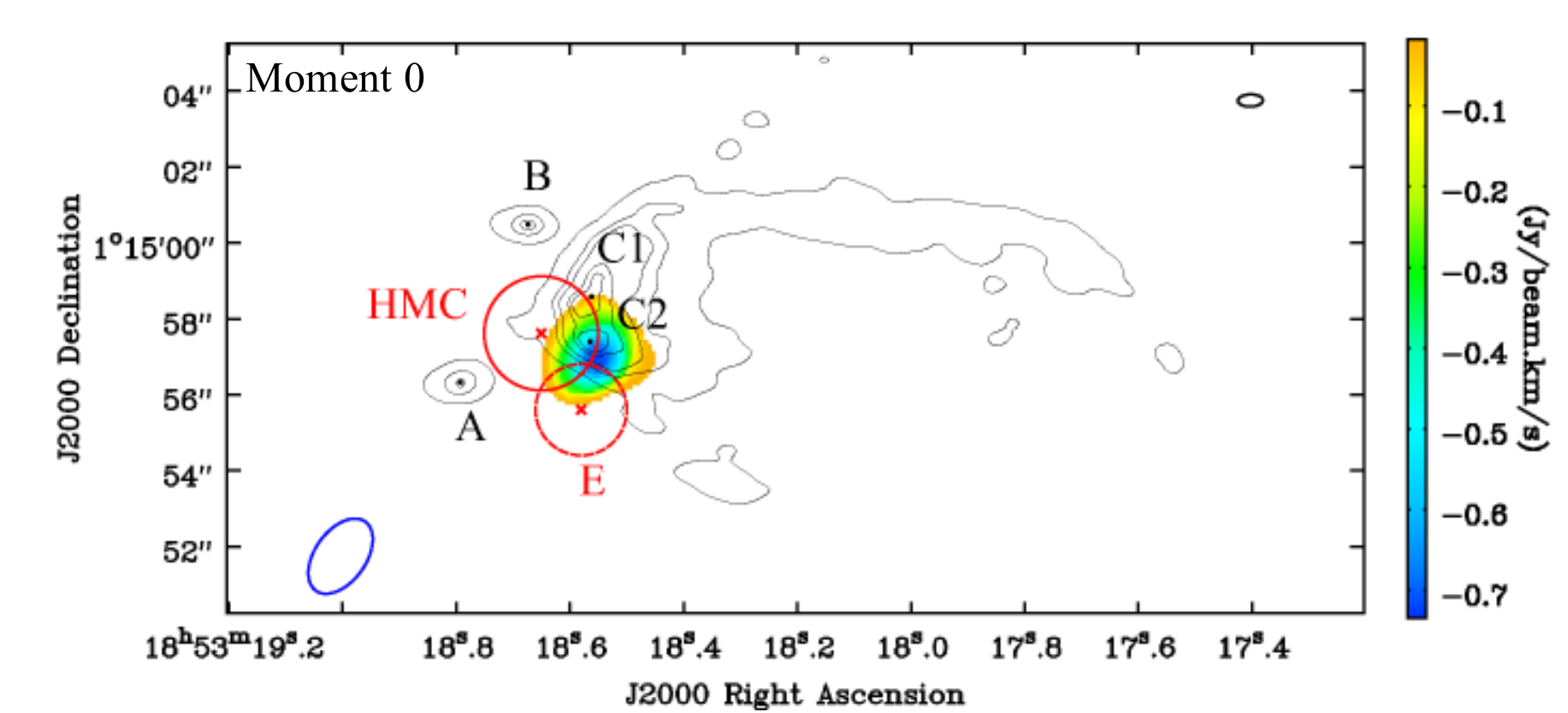}\\
\includegraphics[trim={0 0cm 0 0 },width=0.8\textwidth]{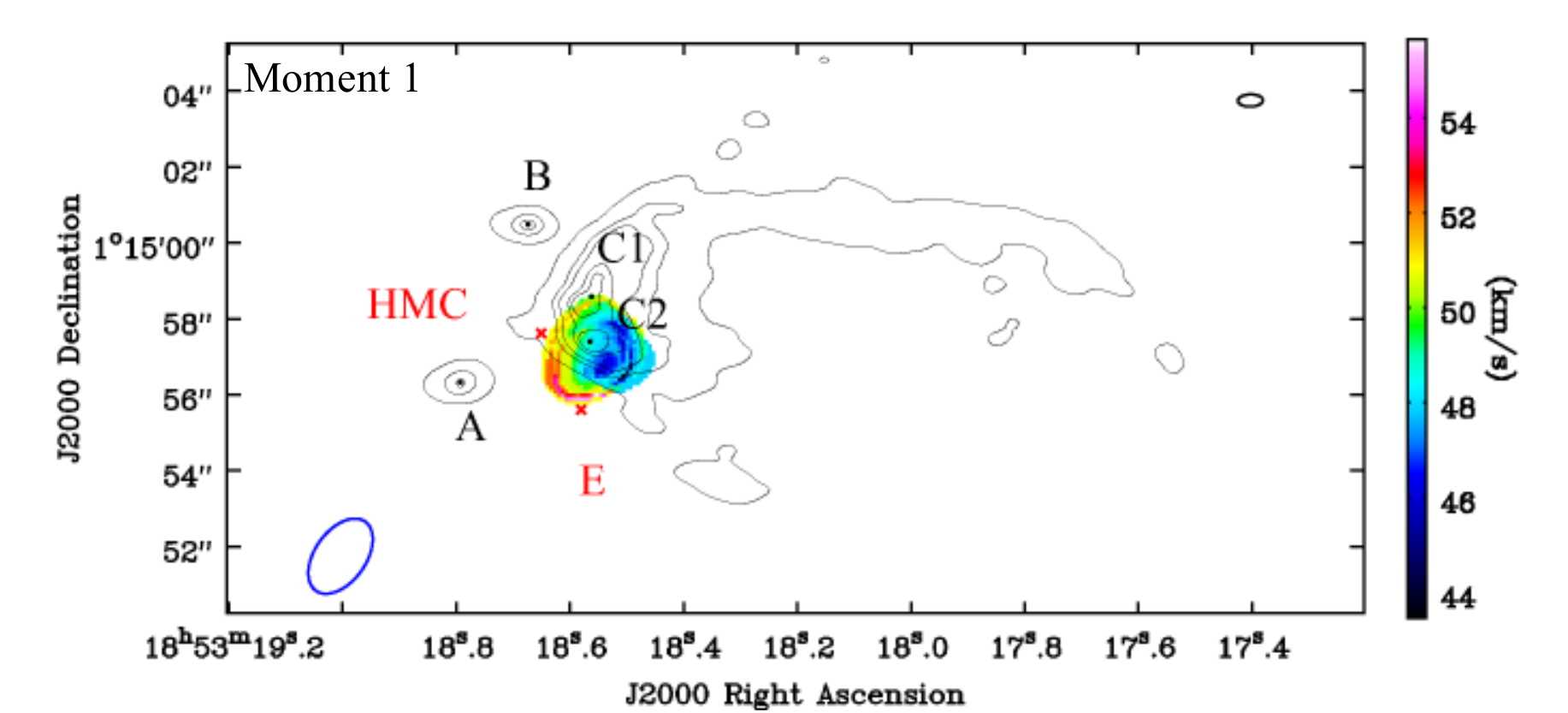}\\
\includegraphics[trim={0 0cm 0 0 },width=0.8\textwidth]{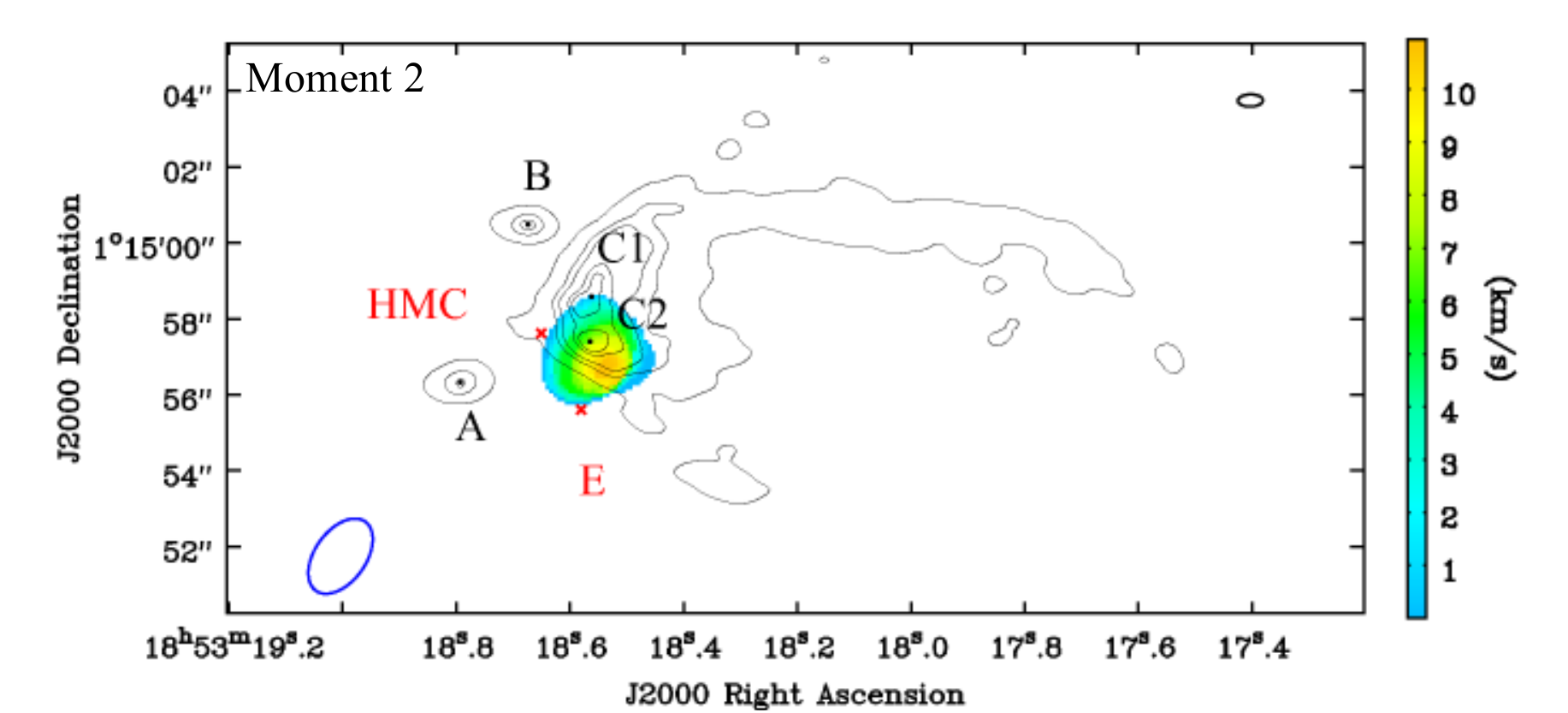}\\
\caption{Radio continuum image of the G34.26+0.15 HCH{\small II} regions (A and B) and the cometary UCH{\small II} region with embedded continuum cores C1 and C2 are shown in contours at 8 GHz in the three panels (image from the VLA image archive, https://www.vla.nrao.edu/astro/nvas/; contours are {0.01, 0.18, 0.37, 0.69, 0.83 $\times\,$0.109$\,$Jy$\,$b$^{-1}$}; synthesized beam in shown in the top right corner). The top panel shows the 6035$\,$MHz OH integrated intensity (moment 0) of the absorption obtained with the VLA in B-configuration (colors); the middle and lower panels show in colors the velocity (moment 1) and velocity dispersion (moment 2) images, respectively. The HCH{\small II} regions A and B, and the two continuum cores of the UCH{\small II} region (C1 and C2) are labeled in all panels. We also show the location of the hot molecular core (HMC) and far infrared source E from \cite{Heaton_1989A&A...213..148H} and \cite{Campbell2000ApJ...536..816C}, respectively (the top panel shows the approximate angular size of the HMC and infrared source E). We note that at a distance of 3.0$\,$kpc, 1\arcsec~ is equivalent to 0.015$\,$pc or 3000$\,$au. 
The synthesized beam of the VLA-B moment images is shown in the lower-left corner.}
\label{fig:8GHz_contour_cont}
\end{figure}
\end{center}

The radio continuum of the G34.26+0.15 complex has been extensively studied in the literature (e.g., \citealt{Reid_1985ApJ...288L..17R}, \citealt{Wood_and_Churchwell_1989ApJS...69..831W}, \citealt{Fey_1994ApJ...435..738F}, \citealt{Sewilo_HII2011ApJS..194...44S}). This site of high-mass
star formation contains a compact H{\small II} region with a diameter of $\sim 1.5$\arcmin~(see continuum images from the VLA D-configuration observations shown in the center column of Figure~\ref{fig:AO-VLA_Spectra_and_Images}), which is known as source D (e.g., \citealt{Fey_1994ApJ...435..738F}, \citealt{Gomez_2000RMxAA..36..161G}, \citealt{Campbell2004ApJ...600..254C}). At the VLA D-configuration resolution, a bright (almost unresolved) continuum source is seen toward the North-West of source D. Arcsecond resolution of the bright continuum source reveals a cometary UCH{\small II} region with a diffuse tail propagating toward the West (source C) and two HCH{\small II} regions (sources A and B) to the East of the UCH{\small II} region.\footnote{The nomenclature used here is consistent with \cite{Sewilo_HII2011ApJS..194...44S}, however, the cometary UCHII region has also been labeled as `A' in the literature \citep{Wood_and_Churchwell_1989ApJS...69..831W}.}
In Figure~\ref{fig:8GHz_contour_cont}, we show in contours an 8$\,$GHz radio continuum sub-arcsecond angular resolution image of the cometary UCH{\small II} and HCH{\small II} regions from the VLA image archive (https://www.vla.nrao.edu/astro/nvas/), where we label the different continuum sources. As previously mentioned in the literature (e.g., \citealt{Sewilo_2004ApJ...605..285S}), the cometary UCH{\small II} region has two main continuum cores, known as C1 and C2. It is unclear whether C1 and C2 are excited by different stars or are simply density and/or temperature enhancements in the larger cometary H{\small II} region.

G34.26+0.15 also harbors a HMC towards the eastern edge of the cometary UCH{\small II} region (e.g., \citealt{Heaton_1989A&A...213..148H}; \citealt{Mookerjea_Kinematics_and_chemistry_2007ApJ...659..447M}). \cite{Campbell2000ApJ...536..816C} reported mid-infrared counterparts of the radio sources A, C, and D (see Figures~\ref{fig:AO-VLA_Spectra_and_Images} and \ref{fig:8GHz_contour_cont}), no counterpart at the location of source B, and an additional source (E) toward the South of C (we show the location and approximate angular size of the 20$\mu m$ source E in Figure~\ref{fig:8GHz_contour_cont}). We also show in Figure~\ref{fig:8GHz_contour_cont} the integrated intensity (moment 0), velocity field (moment 1) and velocity dispersion (moment 2) of the 6035$\,$MHz OH absorption in the region. 

We found that the broad OH absorption is closer to the C2 core. The velocity field image suggests a slight East-West velocity gradient, and comparing the velocity field with the moment 2 image, we note a greater velocity dispersion toward the most blue-shifted core near the center of the OH absorption (slightly South of C2, see Figure~\ref{fig:8GHz_contour_cont} middle and lower panels). However, we stress that the velocity structure of the OH absorption suggested by the moment 1 and 2 images should be considered as tentative, as the moment 0 absorption is unresolved. The coincidence of the 6.0$\,$GHz OH absorption with C2 shows that the absorption is not directly related to the ground state (18$\,$cm) OH masers in the region, which appear to trace the interface between the cometary H{\small II} region and the HMC, as well as the HCH{\small II} region B (e.g., see Figure~3 of \citealt{Zheng_2000MNRAS.317..192Z}).

It is possible that the OH absorption is tracing an outflow that originates from a source offset from the core of the cometary H{\small II} region. For instance, the OH absorption could be tracing an outflow that originates from either HCH{\small II} region to the East of the cometary H{\small II} region, given that \cite{Gaume_1994ApJ...432..648G} proposed that stellar winds or outflows from the HCH{\small II} regions A or B could be the responsible for the cometary morphology of G34.26+0.15C. \cite{Hatchell_2001A&A...372..281H} also pointed out that an SiO blueshifted outflow detected to the North-West of G34.26+0.15C could be driven by G34.26+0.15B. Alternatively, 
as shown in Figure~\ref{fig:8GHz_contour_cont}, the C2 region is partially blended with both the source E (to the South) and the HMC (to the East); the outflow could originate from either source. We note that the arc of H$_2$O masers near the HMC (\citealt{Gomez_2000RMxAA..36..161G}, \citealt{Fey_1994ApJ...435..738F}) shows a blue-shifted North-South velocity gradient that extends to $\sim 0$\kms. However, as shown by the VLBI observations of the H$_2$O masers by \cite{Imai_2011PASJ...63.1293I} (see their Figure~5), the water masers are distributed to the East of the cometary H{\small II} region, coincident with the HMC and not with C2. Therefore, it is unclear whether there is a connection between the H$_2$O velocity gradient and the OH absorption\footnote{However, \cite{Gaume_1994ApJ...432..648G} detected some H$_2$O masers at the location of C2 (see their Figure~6).}. In addition, \cite{Mookerjea_Kinematics_and_chemistry_2007ApJ...659..447M} found evidence that the HMC is externally heated by the cometary H{\small II} region instead of internally heated, therefore, it is unclear whether a young stellar object could be driving an outflow from the HMC. We note, however, that \cite{Keto_1992ApJ...401L.113K} found evidence for low-mass star formation at the HMC location. 

An argument against the OH absorption tracing a molecular outflow originating from the HMC (or from source E, or the HCH{\small II} regions) is the lack of spatial elongation of the absorption profile. If the outflow originates from the HMC, in a way that the blue-shifted cone of the flow happens to be projected toward the cometary H{\small II} region, then one would expect to see the absorption tracing a filament-like structure between the HMC and C2 and possibly continuing beyond C2 (likewise, one would expect to see a North-South elongation of the absorption if an inclined outflow were to originate from source E). This suggests that the OH expanding gas is unlikely to be tracing an inclined outflow originating from the HMC, source E, or the HCH{\small II} regions.

In contrast, the spatial coincidence between the OH absorption and C2 suggests a direct association. This hypothesis is further supported by the kinematics of the ionized gas. As discussed by \cite{Sewilo_HII2011ApJS..194...44S}, the linewidth of the H53$\alpha$ RRL toward the source C is unusually broad, with a full width at zero intensity that ranges from $\sim 0$ to 100\kms, with a line center at $V_{LSR} =  49.9$\kms. The higher angular resolution observations of the H76$\alpha$ line by \cite{Sewilo_2004ApJ...605..285S} were able to resolve the RRL emission from C1 and C2. The C1 profile seems to be caused by the overlap of two different velocity components, while the C2 H76$\alpha$ line has a more symmetric (single Gaussian with center at 47.4\kms) profile, with a full width at zero intensity that also extends to $\sim 0$\kms~(see also \citealt{Garay_1986ApJ...309..553G}). Thus, the complete blue-shifted line-wing of the OH absorption reported here is kinematically and spatially coincident with the blue-shifted side of the RRL line toward C2. 

The association between the broad OH absorption and C2 supports the interpretation that broad radio recombination lines are due to kinematics (turbulence and/or flows) of the plasma, instead of due to pressure broadening as also concluded by \cite{Garay_1986ApJ...309..553G} (see also \citealt{Kurtz_2005IAUS..227..111K}), however, the reason why the OH absorption is only detected toward C2 is unclear. OH is a common molecule in the ISM, if the OH gas in the G34.26+0.15 complex is due to molecular dissociation in the interface between the molecular envelope and the ionized gas, then it is unclear why absorption is not detected toward C1, which has as much continuum as C2, and therefore, would show a similar absorption profile for the same OH opacity. Of course, it is possible that, by chance, there is a molecular core with OH overabundance that happens to be in front of C2, and that is somehow entrained in an ionized flow from the H{\small II} region as revealed by the broad RRLs, e.g., a proplyd-like core caught in the ionized flow. However, such scenario seems to be highly fortuitous. 

\begin{center}
\begin{figure}
\centering
\includegraphics[angle = 270, width=\textwidth]{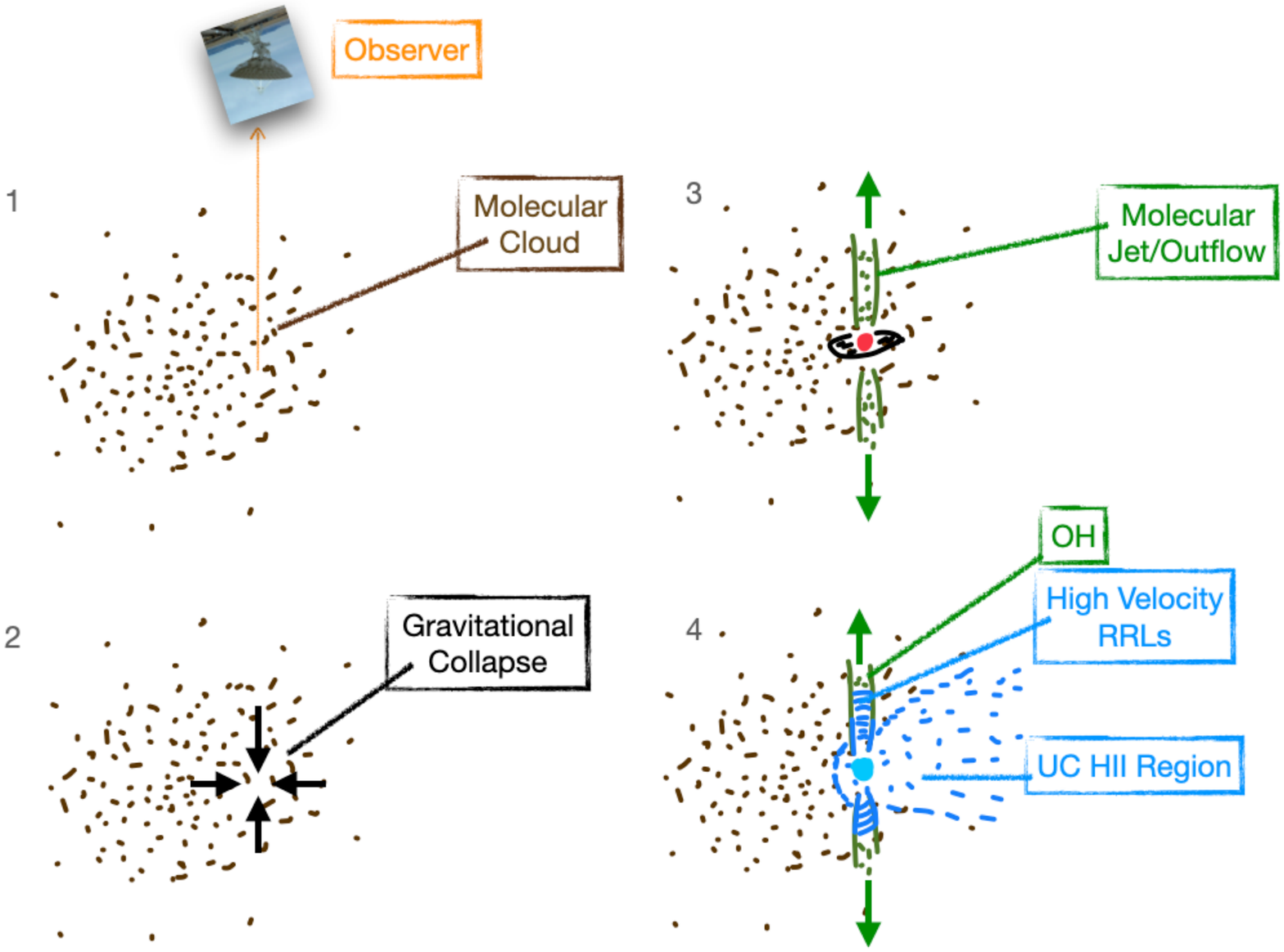}
\caption{Sketch of a possible origin of the high-velocity OH absorption. All panels show a side-view representation of the evolution of the cometary H{\small II} region in G34.26+0.15, in all panels the observer is looking at the star forming region from above (i.e., pole-on in case of panels 3 and 4). Star formation occurs in a Giant Molecular Cloud (panel 1) due to gravitational collapse (panel 2). A protostar and accretion disk form, leading to high-velocity molecular outflows/jets (panel 3). High energy UV photons ionize the molecular gas creating an UCH{\small II} region, the ionization front propagates through the old molecular outflow resulting in high-velocity RRLs, the remnant of the outflow is also detected by OH absorption (panel 4). We emphasize that this is a highly simplified sketch, intended to illustrate how an ionization front might overtake a pre-existing molecular outflow in the case of pole-on observations. The orientation of outflow shown here is not intended to describe the particular geometry of the flow with respect to the HCH{\small II} regions, the HMC or the ionized cores C1 and C2.}
\label{fig:outflow_model}
\end{figure}
\end{center}

We propose an alternative hypothesis of the origin of the compact OH absorption, in which the absorption is revealing a more general feature of high-mass star formation instead of a curious peculiarity of a unique source. Our idea is illustrated in Figure~\ref{fig:outflow_model}. In the canonical model of high-mass star formation via core accretion, star formation occurs in Giant Molecular Clouds where gravitational instabilities lead to collapse, the development of a protostar with an accretion disk and the generation of a molecular jet/outflow. Once the protostar reaches the Zero Age Main Sequence, high-energy UV photons ionize the molecular environment leading to the formation of an H{\small II} region. We propose that the ionization front can propagate through the molecular outflow (explaining the high-velocity RRLs) and near the ionizing front within the outflow, simple molecules like OH can be present traveling at the high speeds of the flow. In the case of G34.26+0.15 C2, we happen to be observing the outflow pole-on, therefore, only the foreground blue-shifted OH is detected in absorption against the continuum of the UC HII region and ionized outflow. We note that this interpretation does not preclude that C1 is ionized by the same massive young stellar object as C2.

This hypothesis would explain why, as observed, the OH absorption has a small angular size and large blue-shifted velocity dispersion due to the pole-on orientation. Given that outflows seem to be less collimated for {\it more evolved} and/or massive objects (see discussion in \citealt{Beuther_and_Shepherd_2005ASSL..324..105B}), it would be possible for the high velocity RRLs to have a larger angular extent (less collimated, more evolved ionized flow) than the more collimated remnant molecular outflow traced by OH (i.e., a `fossil' signature of a less evolved phase). We note that \cite{Garay_1986ApJ...309..553G} argued that the broad RRLs in G34.26+0.15C could be caused by micro-turbulent motion of dense ionized gas expanding towards a low-density medium; the difference with the idea proposed here is that the RRLs are also tracing the ionized flow along a previous molecular outflow, rather than only the champagne expansion of the H{\small II} region. The scenario proposed here is directly related to the idea of outflow-confined H{\small II} regions, where the outflow cavity begins to be ionized while the infalling molecular envelope can confine the rest of the H{\small II} region; the ionized flow results in broad RRLs (\citealt{Tanaka_2016ApJ...818...52T}, \citealt{Tan_2014prpl.conf..149T}, \citealt{Tan_2003astro.ph..9139T}). 

Higher angular resolution observations of the OH absorption and RRL velocity distribution are needed to test this hypothesis, but if confirmed, the combination of broad RRLs and OH absorption could give us a view into the inside-out destruction of molecular outflows by ionized flows in the final phases of high-mass star formation. Higher angular resolution observations will also help to test alternative hypotheses, such as whether the OH absorption traces the destruction of a proplyd-like molecular core from a low mass stellar object at the edge of the H{\small II} region or whether the absorption traces a narrow molecular jet whose blue-shifted axis happens to be in front of the H{\small II} region, but not directly connected to an ionized flow. In addition, high-angular resolution observations of mm RRLs like H30$\alpha$ are also needed. Following the argument about the optical depth effects mentioned in \cite{Sewilo_2008ApJ...681..350S} (see their Sect. 3.3), the lower quantum numbers would trace deeper, denser material, not so much affected by the outer RRL gas layers that are mainly traced by quantum numbers $>$50. Optical depth effects in which RRLs from different energy levels trace different ionized regions (e.g., \citealt{Sams_1996ApJ...459..632S}) can result in biases in the peak RRL velocity used to assess the kinematics of molecular absorption. Therefore, a detailed analysis of RRL emission toward the OH absorption location is required. The need for high-angular resolution observations to evaluate the interpretation of molecular absorption is also exemplified by the single-dish detection of an inverse P-Cygni NH$_3$ profile toward W3(OH), which instead of tracing infalling material, was caused by different NH$_3$ clouds in emission and absorption (e.g., \citealt{Wilson_1978A&A....63....1W}, \citealt{Reid_1987ApJ...312..830R}, \citealt{Wilson_1993ApJ...402..230W}).

\subsection{The Nature of the Excited OH Gas}\label{Analysis_and_Disc:OH}

The detection and non-detection of multiple OH transitions reported in this article (Figures~\ref{fig:Arecibo_spectra} and \ref{fig:AO-VLA_Spectra_and_Images}) allow us to investigate the physical conditions needed to explain the observed spectra. We use the non-LTE radiative transfer and line excitation code {\tt MOLPOP-CEP} \citep{Molpop_2018AA...616..A131} to model our multi-transition data assuming homogeneous conditions. As a starting point to set the basic parameters, we used {\tt MOLPOP-CEP} to reproduce the model of 3$\,$GHz CH emission reported by \cite{Jacob_CH_2021AA...650..A133}, which included analysis of G34.26+0.15 observations. Specifically, we were able to reproduce the results of their Figure~A.1, including $T_{kin} = 75\,$K, density $n \sim 10^3\,$cm$^{-3}$, CH linewidth of 6\kms, Cosmic Microwave Background (CMB), an interstellar radiation field (ISRF) with optically thick dust at a temperature of 40$\,$K ($\tau = 2$), CH abundance $\chi_{CH} =  [CH]/[H_2] = 3.5\times 10^{-8}$ (A. M. Jacob, private communication). We then modified the parameters to model the OH lines, in particular, we explored a grid of densities from n$_{H_{2}}$ = $10^3$ to $10^6$ cm$^{-3}$ with a step of 5$\times10^4$ cm$^{-3}$, linewidths between 1$\,$kms$^{-1}$ to 60$\,$kms$^{-1}$ with a steps of 3$\,$kms$^{-1}$, and kinetic temperatures between 30$\,$K and 300$\,$K (in steps of 15$\,$K). We used the OH collision rate coefficients reported by \cite{Offer_OH_collisions_doi:10.1063/1.466950} and an ortho-to-para H$_{2}$ ratio of 3:1 (e.g., see \citealt{Reeves_1979ZNatA..34..163R}, but no significant difference in our models was obtained when using a 1:1 ortho-to-para ratio).

\begin{center}
\begin{figure}
\centering
\includegraphics[width=\textwidth]{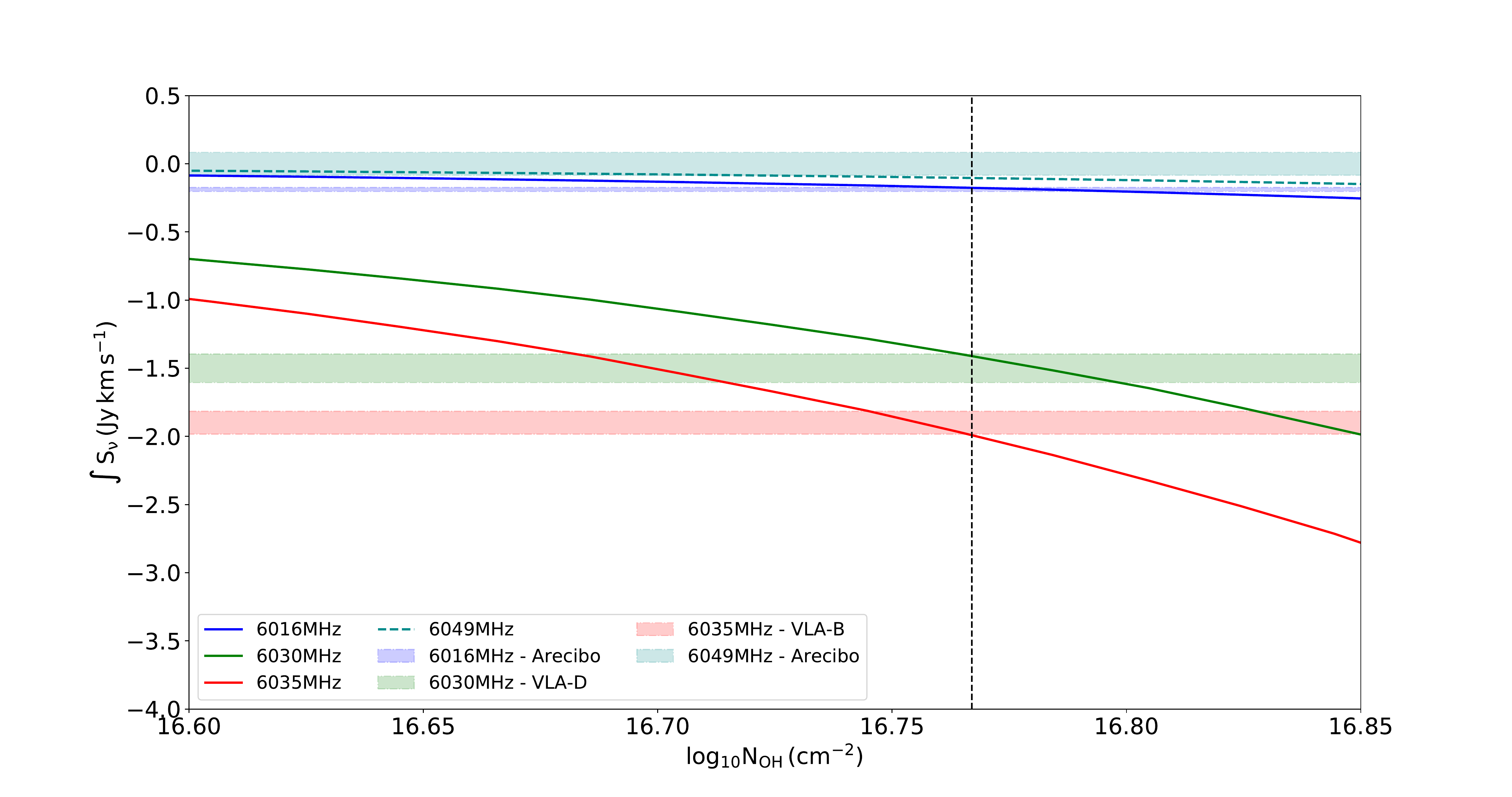}
\caption{Predicted integrated flux density of the \textcolor{blue}{6016$\,$MHz} (blue solid line), 
\textcolor{DrakGreen}{6030$\,$MHz} (green solid line), \textcolor{red}{6035$\,$MHz} (red solid line), and 
\textcolor{LimeBlue}{6049$\,$MHz} (dark-cyan dashed line) OH transitions as a function of column density 
computed with {\tt MOLPOP-CEP} (model assumes T$_{k} = 150\, $K, $n_{H_2} = 10^{5}\,$cm$^{-3}$, 
$~\chi_{OH} = {5\times10^{-6}}$, FWHM = 33$\,$km$\,$s$^{-1}$, FWHM$_{6016MHz}$ = 17$\,$km$\,$s$^{-1}$). The horizontal bands show selected integrated flux density values (including uncertainties) as listed in 
Table~\ref{tab:OH_Line_Parameters}. The vertical dashed line shows a column density for which the model 
matches the observations.
}
\label{fig:oh_Line_parameters}
\end{figure}
\end{center}

An example of the results of the model is presented in Figure~\ref{fig:oh_Line_parameters}, where we show the predicted velocity-integrated flux density of the lines as a function of OH column density (solid and dashed thick lines; the figure includes only 6.0$\,$GHz OH transitions).  Figure~\ref{fig:oh_Line_parameters} also shows, with horizontal bands, the integrated flux density measurement of the lines (or best upper limit) as reported in Table~\ref{tab:OH_Line_Parameters} (the 6016$\,$MHz and 6049$\,$MHz OH measurements are from Arecibo, the 6030$\,$MHz is from the VLA D-configuration data, and the 6035$\,$MHz is from the VLA B-configuration data). The parameters used to generate the model were: $n_{H_2} = 10^5\,$cm$^{-3}$, $T_{kin} = 150\,$K, OH abundance $\chi_{OH} = 5 \times 10^{-6}$, a full width half maximum $FWHM = 17$\kms~for the 6016$\,$MHz line and $FWHM = 33$\kms~for the other lines, and background continuum from the UCH{\small II} region with a brightness temperature $T_{B, cont} = 3000\,$K (obtained from the background continuum of the 6035$\,$MHz absorption from the VLA-B observations shown in Figure~\ref{fig:AO-VLA_Spectra_and_Images}, right panels).
As shown in Figure~\ref{fig:oh_Line_parameters}, this set of parameters reproduces (within the quoted uncertainties) the observed OH absorption and non-detection for a column density ($N_{OH}$) slightly below $10^{16.8}\,$cm$^{-2}$. We highlight that we used different linewidths to reproduce the observed integrated flux densities, which suggests that the weak 6016$\,$MHz OH absorption may not be tracing the same gas detected in 6030$\,$MHz and 6035$\,$MHz OH absorption. The 6016$\,$MHz OH line was not imaged with the VLA, hence, we do not know if it originates from the same projected direction. The Arecibo spectra had to be significantly smoothed to detect the line, and the spectral line is narrower than the 6030 and 6035$\,$MHz OH absorption (Figure~\ref{fig:Arecibo_spectra}), therefore, assuming that the 6016$\,$MHz OH line is tracing slightly different material  is consistent with the available data. We note that multiple velocity components were used by \cite{Hajigholi_accretion_process_2016A&A...585A.158H} to model the infall velocity structure of G34.26+0.15, therefore, using two different linewidths to model the OH absorption is reasonable. Moreover, we know that different transitions of OH can trace significantly different environments in G34.26+0.15 as the 1665 and 1667$\,$MHz OH masers trace the eastern edge of the cometary UCH{\small II} region; the masers are found in the arc between the ionized gas and the HMC (as well as associated with the HCH{\small II} region B), and no maser is found toward C2 \citep{Zheng_2000MNRAS.317..192Z}. Similarly, the $^2\Pi_{1/2}~ J = 3/2 - 1/2$ rotational OH transitions observed with SOFIA, i.e., the transitions connecting the 7.8$\,$GHz ladder (not-observed in this work) and the 4.7$\,$GHz ladder (detected in this work, Figure~\ref{fig:Arecibo_spectra}) show emission at the systemic velocity with red-shifted self-absorption, and no indication of the blue-shifted gas reported here \citep{Csengeri_2022A&A...658A.193C}.

Despite the simplistic assumptions used to generate Figure~\ref{fig:oh_Line_parameters}, our model shows that reasonable physical conditions can explain the detections and non-detections of the 6.0$\,$GHz OH lines. For instance, a kinematic temperature of 150$\,$K is consistent with the temperature of molecular gas externally heated by the UCH{\small II} region C \citep{Mookerjea_Kinematics_and_chemistry_2007ApJ...659..447M}. The density $n_{H_2} = 10^5\,$cm$^{-3}$ suggests that the OH is tracing a high-density component of the outflow (similar to CH in outflows as discussed by \citealt{Magnani_1992A&AS...93..509M}). We note that  temperatures of $\sim 100\,$K and density values $n_{H_2} \sim 10^5 - 10^6\,$cm$^{-3}$ are also the conditions suggested by SiO observations of the outflow in the high-mass star forming region G5.89$-$0.39 \citep{Acord_1997ApJ...475..693A}. Such densities in outflows are also reasonable based on theoretical models (e.g., \citealt{Tanaka_2016ApJ...818...52T}). Likewise, while the OH column density consistent with the observed absorption ($\sim 10^{16} - 10^{17}\,$cm$^{-2}$) is greater than typical values reported for H{\small II} regions \citep{Rugel_thor_abundance_2018A&A...618A.159R}, such a high value has been obtained toward some high-mass star forming regions based on excited OH absorption measurements \citep{Baudry_2002A&A...394..107B}. 

The parameters used to generate Figure~\ref{fig:oh_Line_parameters} can also be compared to the multi-transition modeling of several OH absorption lines (including the 6.0$\,$GHz lines) detected toward the star forming regions DR21 and K3-50 by \cite{Jones_1992A&A...264..220J}. They found that the observed absorption profiles could be explained by molecular gas with $T_k$ between $\sim$100 and 180$\,$K, $T_d < 90\,$K, OH column densities between $10^{15}$ and a few times $10^{16}\,$cm$^{-2}$, molecular hydrogen densities between $10^6$ and $10^8\,$cm$^{-3}$, and OH abundance between $10^{-6}$ and $10^{-7}$. This range of physical conditions is similar to the parameters used to generate Figure~\ref{fig:oh_Line_parameters}\footnote{We note that \cite{Jones_1994A&A...288..581J} revisited the models presented in \cite{Jones_1992A&A...264..220J}, and reported greater $T_d$ and slightly greater $T_k$ values.}. As pointed out by \cite{Jones_1992A&A...264..220J} and \cite{Jones_1994A&A...288..581J}, there is a degeneracy between OH abundance and molecular density. If the OH gas detected here is tracing higher density gas than $10^5\,$cm$^{-3}$, the abundance value is expected to be smaller than $5\times10^{-6}$.

Finally, we note that even though the model can reproduce the observed absorption of the 6.0$\,$GHz lines, it does not predict emission of the 4.7$\,$GHz OH transitions detected with Arecibo (Figure~\ref{fig:Arecibo_spectra}). However, we found that at higher densities ($\sim 10^{8}\,$cm$^{-3}$) and slightly higher temperatures ($\sim 170\,$K), the model predicts absorption of the 6.0$\,$GHz transitions while population inversion is expected for the 4660$\,$MHz OH line at a column density of $\sim 10^{15.8}\,$cm$^{-2}$, and FWHM = 17$\,$km$\,$s$^{-1}$. We note that the model of \cite{Jones_1994A&A...288..581J} (see also  \citealt{Jones_1992A&A...264..220J}) for higher density molecular gas than the one used in Figure~\ref{fig:oh_Line_parameters}, predicted simultaneous absorption from the 6.0$\,$GHz lines and emission (inverted populations) from the 4.7$\,$GHz lines. We conclude that LTE conditions cannot explain the observed line ratios. At this point, however, further modeling of the lines detected with Arecibo (Figure~\ref{fig:Arecibo_spectra}) would be premature, as additional observations with the VLA are needed to obtain the location of the 4.7$\,$GHz OH emission to establish whether a single model should be able to explain all spectral profiles, or whether the 4.7$\,$GHz OH lines are tracing a different volume of gas.

\section{Summary}\label{Summary}

We report multi-transition observations of the 4.7$\,$GHz and 6.0$\,$GHz OH lines toward the high-mass star forming region G34.26+0.15. The 4.7$\,$GHz observations were obtained with the 305$\,$m Arecibo Telescope, while the 6.0$\,$GHz OH lines were observed with Arecibo as well as with the VLA in two configurations, D and B arrays, resulting in a maximum angular resolution of $\sim 2$\arcsec. We detected emission from the 4750$\,$MHz and 4765$\,$MHz OH lines, and tentative detection of 4660$\,$MHz OH emission. In contrast, the 6030$\,$MHz and 6035$\,$MHz lines show broad OH absorption, with a linewidth at zero intensity that expands over 50\kms. The absorption spectra obtained with Arecibo and the VLA are remarkably similar, which shows that the OH is compact. Indeed, the OH 6035$\,$MHz OH absorption was unresolved at the resolution of the VLA B-configuration data. The absorption is found toward the ionized core C2 of the cometary UCH{\small II} region. This core of ionized gas is known to show broad RRL emission; the blue-shifted profile of the RRL line covers the velocity range of the OH absorption, which strongly suggests that the broad RRL is tracing a plasma flow that also involves a molecular outflow. We highlight the possibility that the broad RRL could be tracing the inside-out ionization of a remnant molecular outflow, which is detected in OH absorption seen pole-on. If this interpretation is correct, the combination of broad RRLs and high-velocity excited OH absorption could be indicating the final phase in the evolution of molecular outflows as they are destroyed by ionized flows from H{\small II} regions, i.e., these could be useful tracers to probe the final stages of high-mass star formation. We note that other high-mass star forming regions like DR21 \citep{Guilloteau_1984A&A...131...45G} and G45.07+0.13 \citep{Tan_arecibo_10.1093/mnras/staa1841} also show blue-shifted 6.0 GHz OH linewings (albeit less extreme than the case of G34.26+0.15 discussed here), and therefore, excited OH absorption could be used to investigate the interaction between ionized flows and expanding molecular gas in other high-mass star forming regions. 

We used the non-LTE radiative transfer code {\tt MOLPOP-CEP} to model the observed lines. While we were able to find a set of physical conditions that can explain the observed 6030 and 6035$\,$MHz OH absorption features (and non-detection of the 6049$\,$MHz line), a different velocity dispersion was needed to account for a weak 6016$\,$MHz absorption feature, and the model did not predict emission of the 4.7$\,$GHz lines. The 6016$\,$MHz OH line is significantly weaker than the 6030 and 6035$\,$MHz OH lines; the 6016$\,$MHz OH line was not detected with the VLA and the Arecibo spectrum had to be significantly smoothed to reveal the absorption. High angular resolution and sensitivity observations of the 6016$\,$MHz and 4.7$\,$GHz OH detections are needed to investigate whether the transitions are tracing the same material as the 6030 and 6035$\,$MHz OH absorption lines. Such follow-up observations are needed to reliably explore whether the OH absorption lines reported in this work are consistent with the inside-out ionization of a remnant molecular outflow. We highlight that detectability of the weak spectral lines reported in this work requires large collecting areas; this work shows the potential of high-sensitivity excited OH studies that could be conducted toward a large sample of sources using future instrumentation, such as the ngVLA \citep{Carilli_2015arXiv151006438C} and a possible Next Generation Arecibo Telescope \citep{Roshi_2021arXiv210301367A}.


\acknowledgments
We thank the review and comments from an anonymous referee that helped us improve this manuscript. E.D.A. acknowledges partial support from NSF grant AST-1814063 and computational resources donated by the WIU Distinguished Alumnus Frank Rodeffer. P.H. acknowledges partial support from NSF grant AST-1814011. This research has made use of NASA’s Astrophysics Data System. The National Radio Astronomy Observatory (NRAO) is a facility of the National Science Foundation operated under cooperative agreement by Associated Universities, Inc. The 305$\,$m Arecibo Telescope was a facility of the National Science Foundation operated under cooperative agreement by the University of Central Florida and in alliance with Universidad Ana G. Mendez, and Yang Enterprises, Inc. The authors would like to specially thank A. M. Jacob for providing guidance to reproduce results of \cite{Jacob_CH_2021AA...650..A133} using {\tt MOLPOP-CEP}. E.D.A. would like to thank the NRAO for providing access to the New Mexico Array Science Center (NMASC) cluster for remote data reduction and analysis, which allowed participation of the high-school student Cade Rigg in this project. Computational resources from NSF ACCESS JetStream2 (Indiana University) under project PHY220136 were also used by Cade Rigg. E.D.A. also acknowledges Ms. Kassie Henry, Ms. Sue Henry and Ms. Sara Guymon for encouraging and guiding Cade Rigg during his middle and high-school science projects.

%

\vspace{5mm}
\facilities{VLA (NRAO), Arecibo (UCF, NSF)}


\software{{\tt CASA} \citep{McMullin_2007ASPC..376..127M}; {\tt MOLPOP-CEP} \citep{Molpop_2018AA...616..A131}}







\bibliography{OH_Outflow_G34_VLA}{}
\bibliographystyle{aasjournal}


\end{document}